# Low Gilbert Damping Constant in Perpendicularly Magnetized W/CoFeB/MgO Films with High Thermal Stability

Dustin M. Lattery[1†], Delin Zhang[2†], Jie Zhu[1], Jian-Ping Wang[2]*, and Xiaojia Wang[1]*

[1]Department of Mechanical Engineering, University of Minnesota, Minneapolis, MN 55455, USA

[2]Department of Electrical and Computer Engineering, University of Minnesota, Minneapolis, MN 55455, USA

[†]These authors contributed equally to this work.

*Corresponding authors: wang4940@umn.edu & jpwang@umn.edu

**Abstract:** Perpendicular magnetic materials with low damping constant and high thermal stability have great potential for realizing high-density, non-volatile, and low-power consumption spintronic devices, which can sustain operation reliability for high processing temperatures. In this work, we study the Gilbert damping constant ($\alpha$) of perpendicularly magnetized W/CoFeB/MgO films with a high perpendicular magnetic anisotropy (PMA) and superb thermal stability. The $\alpha$ of these PMA films annealed at different temperatures ($T_{\text{ann}}$) is determined *via* an all-optical Time-Resolved Magneto-Optical Kerr Effect method. We find that $\alpha$ of these W/CoFeB/MgO PMA films decreases with increasing $T_{\text{ann}}$, reaches a minimum of $\alpha = 0.016$ at $T_{\text{ann}} = 350\ °C$, and then increases to 0.024 after post-annealing at 400 °C. The minimum $\alpha$ observed at 350 °C is rationalized by two competing effects as $T_{\text{ann}}$ becomes higher: the enhanced crystallization of CoFeB and dead-layer growth occurring at the two interfaces of the CoFeB layer. We further demonstrate that $\alpha$ of the 400 °C-annealed W/CoFeB/MgO film is comparable to that of a reference Ta/CoFeB/MgO PMA film annealed at 300 °C, justifying the enhanced thermal stability of the W-seeded CoFeB films.

I. INTRODUCTION

Since the first demonstration of perpendicular magnetic tunnel junctions with perpendicular magnetic anisotropy (PMA) Ta/CoFeB/MgO stacks [1], interfacial PMA materials have been extensively studied as promising candidates for ultra-high-density and low-power consumption spintronic devices, including spin-transfer-torque magnetic random access memory (STT-MRAM) [2,3], electrical-field induced magnetization switching [4-6], and spin-orbit torque (SOT) devices [7-9]. An interfacial PMA stack typically consists of a thin ferromagnetic layer (*e.g.*, CoFeB) sandwiched between a heavy metal layer (*e.g.*, Ta) and an oxide layer (*e.g.*, MgO). The heavy metal layer interface with the ferromagnetic layer is responsible for the spin Hall effect, which is favorable for SOT and skyrmion devices [10,11]. The critical switching current ($J_{c0}$) should be minimized to decrease the power consumption of perpendicular STT-MRAM and SOT devices. Reducing $J_{c0}$ requires the exploration of new materials with low Gilbert damping constant ($\alpha$), large spin Hall angle ($\theta_{SHE}$), and large effective anisotropy ($K_{eff}$) [12,13].

In addition, spintronic devices need to sustain operation reliability for processing temperatures as high as 400 °C for their integration with existing CMOS fabrication technologies, providing the standard back-end-of-line process compatibility [14]. Based on this requirement, the magnetic properties of a PMA material should be thermally stable at annealing temperatures ($T_{ann}$) up to 400 °C. Unfortunately, Ta/CoFeB/MgO PMA films commonly used in spintronic devices cannot survive with $T_{ann}$ higher than 350 °C, due to Ta diffusion or CoFeB oxidation at the interfaces [15,16]. The diffusion of Ta atoms can act as scattering sites to increase the spin-flip probability [17] and lead to a higher Gilbert Damping constant ($\alpha$), a measure of the energy dissipation from the magnetic precession into phonons or magnons [18].

Modifying the composition of thin-film stacks can prevent heavy metal diffusion, which is beneficial to both lowering $\alpha$ and improving thermal stability [19]. Along this line, new interfacial PMA stacks have been developed, such as Mo/CoFeB/MgO, to circumvent the limitation on device processing temperatures [20,21]. While Mo/CoFeB/MgO films can indeed exhibit PMA at temperatures higher than 400 °C, they cannot be used for SOT devices due to the weak spin Hall effect of the Mo layer [20,21]. Recently, W/CoFeB/MgO PMA thin films have been proposed because of their PMA property at high post-annealing temperature [22], and the large spin Hall angle of the W layer ($\theta_{SHE} \approx 0.30$) [23], which is twice that of a Ta layer ($\theta_{SHE} \approx 0.12 \sim 0.15$) [9]. While there have been a few scattered studies demonstrating the promise of fabricating SOT devices using the W/CoFeB/MgO stacks, special attention has been given to their PMA properties and functionalities as SOT devices [24]. A systematic investigation is lacking on the effect of $T_{ann}$ on $\alpha$ of W/CoFeB/MgO PMA thin films.

## II. SAMPLE PREPERATION AND MAGNETIC CHARACTERIZATION

In this work, we grow a series of W(7)/Co$_{20}$Fe$_{60}$B$_{20}$(1.2)/MgO(2)/Ta(3) thin films on Si/SiO$_2$(300) substrates (thickness in nanometers) with a magnetron sputtering system (<5×10$^{−8}$ Torr). These films are then post-annealed at varying temperatures ($T_{ann} = 250 \sim 400$ °C) within a high-vacuum furnace (<1×10$^{−6}$ Torr) and their magnetic properties and damping constants as a function of $T_{ann}$ are systematically investigated. For comparison, a reference sample of Ta(7)/Co$_{20}$Fe$_{60}$B$_{20}$(1.2)/MgO(2)/Ta(3) is also prepared to examine the effect of seeding layer to the damping constant of these PMA films. The saturation magnetization ($M_s$) and anisotropy of these films are measured with the Vibrating Sample Magnetometer (VSM) module of a Physical

Property Measurement System. Figure 1 plots the magnetic hysteresis loops and associated magnetic properties extracted from VSM measurements.

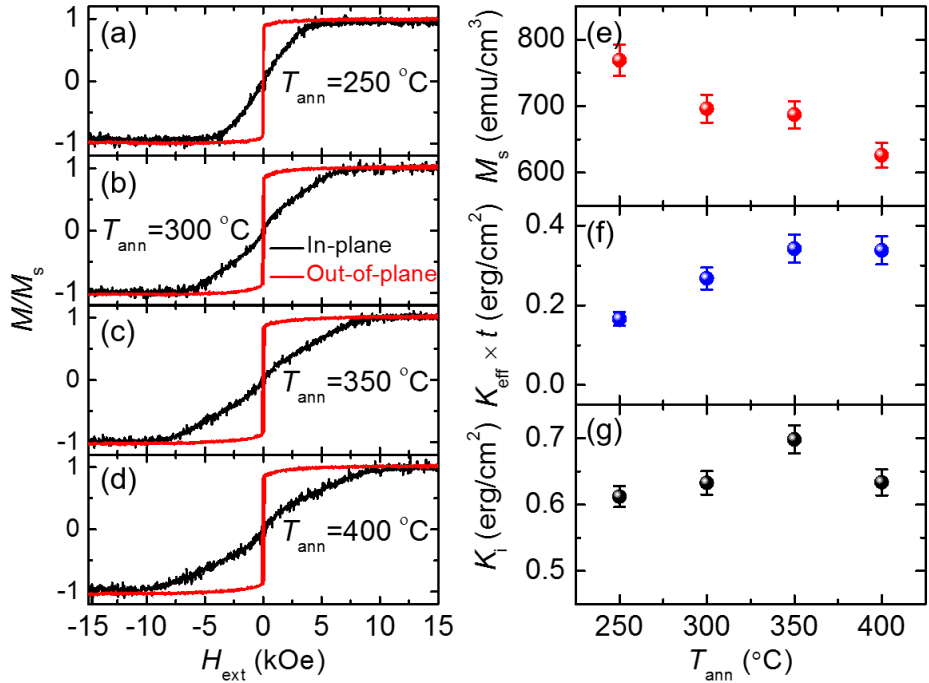

**Figure 1.** Room temperature magnetic hysteresis loops of W/CoFeB/MgO PMA thin films post-annealed at (a) 250 °C, (b) 300 °C, (c) 350 °C, and (d) 400 °C. Black and red curves denote external magnetic field ($H_{ext}$) applied along and perpendicular to the film plane, respectively. (e-g) Plots of the saturation magnetization ($M_s$), effective interfacial anisotropy ($K_{eff} \times t$), and interfacial anisotropy ($K_i$) as functions of $T_{ann}$.

With the increase of $T_{ann}$, $M_s$ for the W/CoFeB/MgO films decreases from ~780 to ~630 emu/cm$^3$ [Fig. 1(e)]. The effective interfacial anisotropy [($K_{eff} \times t$) depicted in Fig. 1(f)] shows an increasing trend with $T_{ann}$ (from ~0.18 to ~0.34 erg/cm$^2$ when $T_{ann}$ increases from 250 to 350 °C) and saturates at $T_{ann}$ = 350 °C. The positive values of $K_{eff} \times t$ suggest that these W/CoFeB/MgO films maintain high PMA properties at elevated temperatures including 400 °C, demonstrating their enhanced thermal stability compared to Ta/CoFeB/MgO films that can only sustain PMA up to 350 °C. Removing the influence of the demagnetization energy from $K_{eff} \times t$ results in the interfacial anisotropy ($K_i$), which changes from 0.6 to 0.7 erg/cm$^2$ with the increase of $T_{ann}$ up to 350 °C and then decreases to ~0.6 erg/cm$^2$ at $T_{ann}$ = 400 °C [Fig. 1(f)]. Details about the determination of $K_i$ and $K_{eff}$ are provided in Section S1 of the Supplemental Material (SM).

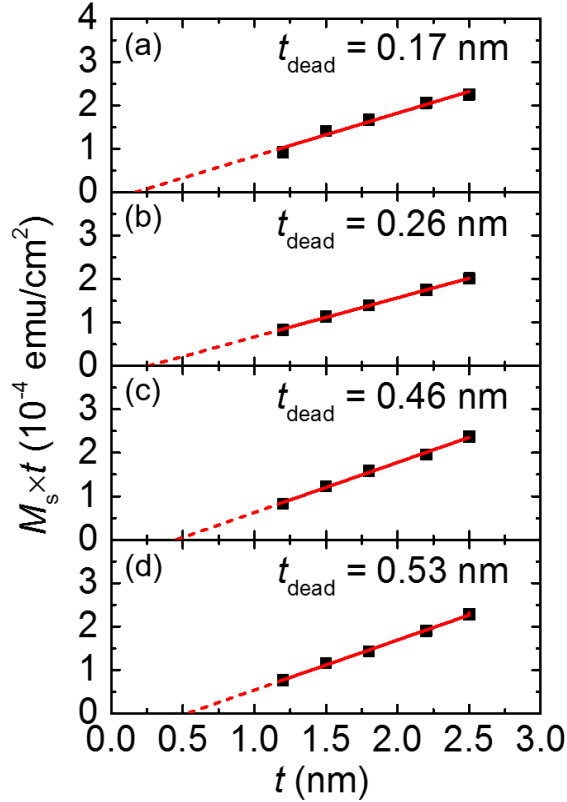

**Figure 2**. The dead-layer extraction results. (a), (b), (c), and (d) represent the series of samples annealed at 250, 300, 350, and 400 °C respectively. The $t_{dead}$ value is the extrapolated $x$-axis intercept from the linear fitting of the thickness-dependent saturation magnetization area product ($M_s \times t$).

We attribute the decrease of $K_i$ at high $T_{ann}$ to the growth of a dead layer at the CoFeB interfaces, which becomes prominent at higher $T_{ann}$. To quantitatively determine the thickness of the dead layer as $T_{ann}$ increases, we prepare four sets of PMA stacks of W(7)/CoFeB($t$)/MgO(2)/Ta(3). One set contains five stacks with varying thicknesses for the CoFeB layer ($t$ = 1.2, 1.5, 1.8, 2.2, and 2.5 nm) and is post-annealed at a fixed $T_{ann}$. Four $T_{ann}$ of 250, 300, 350, and 400 °C are used for four sets of the PMA stacks, respectively. The annealing conditions are the same as those for the W(7)/CoFeB(1.2)/MgO(2)/Ta(3) samples for discussed previously. We measure the magnetic hysteresis loops of these samples using VSM and plot their saturation magnetization area product ($M_s \times t$) as a function of film thickness ($t$) in Fig. 2. Linear extrapolation of the $M_s \times t$ data provides the dead-layer thickness, at which the magnetization reduces to zero as illustrated by the *x*-axis intercept in Fig. 2.

## III. TR-MOKE MEASUREMENTS

The magnetization dynamics of these PMA thin films are determined using the all-optical Time-Resolved Magneto-Optical Kerr Effect (TR-MOKE) method [25-29]. This pump-probe method utilizes ultra-short laser pulses to thermally demagnetize the sample and probe the resulting Kerr rotation angle ($\theta_K$). In the polar-MOKE configuration, $\theta_K$ is proportional to the change of the out-of-plane component of magnetization [$\Delta M_z$ in Fig. 3(a)] [30]. Details of the TR-MOKE setup are provided in Section S2 of the SM.

The TR-MOKE signal is fitted to the equation $\theta_K = A + Be^{-t/C} + D\sin(2\pi ft + \varphi)e^{-t/\tau}$, where *A*, *B*, and *C* are the offset, amplitude, and exponential decaying constant of the thermal background, respectively. *D* denotes the amplitude of oscillations, *f* is the resonance frequency, $\varphi$ is a phase shift (related to the demagnetization process), and $\tau$ is the relaxation time of magnetization precession. Directly from TR-MOKE measurements, an effective damping constant ($\alpha_{eff}$) can be extracted based on the relationship $\alpha_{eff} = 1/(2\pi f \tau)$. However, $\alpha_{eff}$ is not an intrinsic material property; rather, it depends on measurement conditions, such as the applied field direction [$\theta_H$ in Fig. 3(a)], the magnitude of the applied field ($H_{ext}$), and inhomogeneities of the sample (*e.g.* local variation in the magnetic properties of the sample) [31,32].

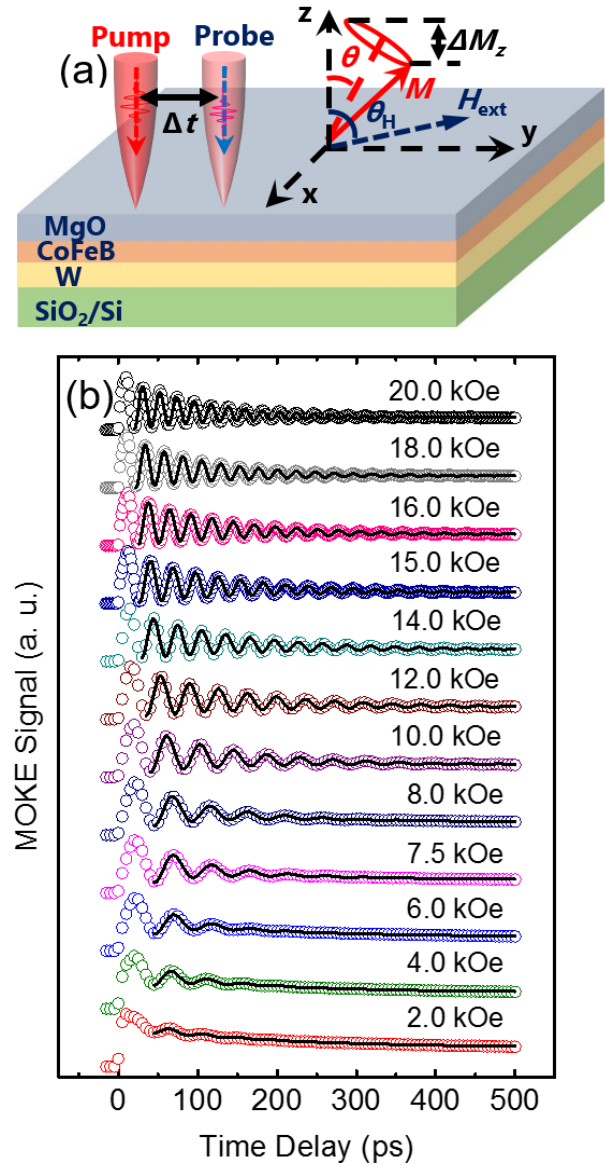

**Figure 3.** (a) Definition of the parameters and angles used in TR-MOKE experiments. The red circle indicates the magnetization precession. $\theta$ is the equilibrium direction of the magnetization. $\theta_K$ is measured by the probe beam at a given time delay ($\Delta t$). (b) The TR-MOKE data (open symbols) and model fitting of $\theta_K$ (black curves) for the 400 °C sample at 76°, for varying $H_{ext}$ from 2.0 to 20 kOe.

To obtain the Gilbert damping constant, the inhomogeneous contribution needs to be removed from $\alpha_{eff}$, such that the remaining value of damping is an intrinsic material property and independent of the measurement conditions. To determine the inhomogeneous broadening in the sample, the effective anisotropy field ($H_{k,eff} = 2K_{eff}/M_s$) needs to be pre-determined from either (1) the magnetic hysteresis loops; or (2) the fitting results of $f$ vs. $H_{ext}$ obtained from TR-MOKE. The resonance frequency, $f$, can be related to $H_{ext}$ through the Smit-Suhl approach by identifying the second derivatives of the total magnetic free energy, which combines a Zeeman energy, an anisotropy energy, and a demagnetization energy [33-35]. For a perpendicularly magnetized thin film, $f$ is defined by Eqs. (1-4) [35].

$$f = \frac{\gamma}{2\pi}\sqrt{H_1 H_2}, \tag{1}$$

$$H_1 = H_{ext}\cos(\theta - \theta_H) + H_{k,eff}\cos^2(\theta), \tag{2}$$

$$H_2 = H_{ext}\cos(\theta - \theta_H) + H_{k,eff}\cos(2\theta), \tag{3}$$

$$2H_{ext}\sin(\theta_H - \theta) = H_{k,eff}\sin(2\theta). \tag{4}$$

This set of equations permits calculation of $f$ with the material gyromagnetic ratio ($\gamma$), $H_{ext}$, $\theta_H$, $H_{k,eff}$, and the angle between the equilibrium magnetization direction and the surface normal [$\theta$, determined by Eq. (4)]. The measured values of $f$ as a function of $H_{ext}$ can be fitted to Eq. (1) by treating $\gamma$ and $H_{k,eff}$ as fitting parameters. To minimize the fitting errors resulting from the inhomogeneous broadening effect that is pronounced at the low fields, we use measured frequencies at high fields ($H_{ext} > 10$ kOe) to determine $H_{k,eff}$.

With a known value of $H_{k,eff}$, the Gilbert damping constant of the sample can be determined through a fitting of the inverse relaxation time ($1/\tau$) to Eq. (5). The two terms of Eq. (5) take into account, respectively, contributions from the intrinsic Gilbert damping of the materials (first term) and inhomogeneous broadening (second term) [31]:

$$\frac{1}{\tau} = \frac{1}{2}\alpha\gamma(H_1 + H_2) + \frac{1}{2}\left|\frac{d\omega}{dH_{k,eff}}\right|\Delta H_{k,eff}, \tag{5}$$

where $H_1$ and $H_2$ are related to the curvature of the magnetic free energy surface as defined by Eqs. (2) and (3) [35,36]. The second term on the right side of Eq. (5) captures the inhomogeneous effect by attributing it to a spatial variation in the magnetic properties ($\Delta H_{k,eff}$), analogous to the linewidth broadening effect in Ferromagnetic Resonance measurements [37]. The magnitude of $d\omega/dH_{k,eff}$ can be calculated once the relationship of $\omega$ vs. $H_{ext}$ is determined with a numerical method. Both $\alpha$ and $\Delta H_{k,eff}$ (the inhomogeneous term related to the amount of spatial variation in $H_{k,eff}$) are determined *via* the fitting of the measured $1/\tau$ based on Eq. (5). In this way, we can uniquely extract the field-independent $\alpha$, as an intrinsic material property, from the effective damping ($\alpha_{eff}$), which is directly obtained from TR-MOKE and dependent on $H_{ext}$.

It should be noted here that the inhomogeneous broadening of the magnetization precession is presumably due to the multi-domain structure of the materials, which becomes negligible in the high-field regime ($H_{ext} \gg H_{k,eff}$) as the magnetization direction of multiple magnetic domains becomes uniform. This is also reflected by the fact that the derivative in the second term of Eq. (5) approaches zero for the high-field regime [38].

## IV. RESULTS AND DISCUSSION

The measurement method is validated by measuring the $T_{ann}$ = 400 °C at multiple angles ($\theta_H$) of the external magnetic field direction. By repeating this measurement at varying $\theta_H$, we can show that $\alpha$ is an intrinsic material property, independent of $\theta_H$. Figure 4(a) plots the resonance frequencies derived from TR-MOKE and model fittings for the 400 °C sample at two field directions ($\theta_H$ = 76° and 89°). For the data acquired at $\theta_H$ = 89°, a minimum $f$ occurs at $H_{ext} \approx H_{k,eff}$. This corresponds to the smallest amplitude of magnetization precession, when the equilibrium direction of the magnetization is aligned with the applied field direction at the magnitude of $H_{k,eff}$ [35]. The dip at this local minimum diminishes when $\theta_H$ decreases, as reflected by the comparison between the red ($\theta_H = 89°$) and blue ($\theta_H = 76°$) lines in Fig. 4(a). With the $H_{k,eff}$ extracted from the fitting of frequency data with $\theta_H = 89°$, we generate the plot of theoretically predicted $f$ vs. $H_{ext}$ [$\theta_H = 76°$ theory, blue line in Fig. 4(a)], which agrees well with experimental data [open squares in Fig. 4(a)].

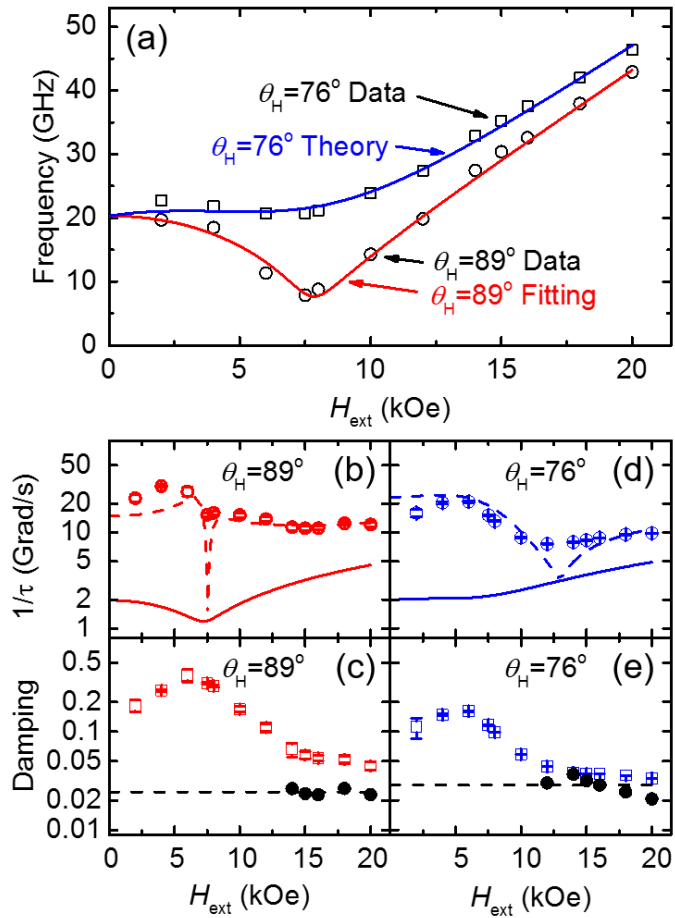

**Figure 4.** (a) Measured $f$ vs. $H_{ext}$ results for the 400 °C sample at $\theta_H = 89°$ (open circles) and $\theta_H = 76°$ (open squares) and corresponding modeling at $\theta_H = 89°$ (red line) and $\theta_H = 76°$ (blue line). (b) The measured inverse of relaxation time ($1/\tau$) at $\theta_H = 89°$ (open symbols) and the fitting of $1/\tau$ based on Eq. (5) (dotted line). For reference, the first term of $1/\tau$ in Eq. (5) is also plotted (solid line), which accounts for the contribution from the Gilbert damping only. (c) $\alpha_{eff}$ as a function of $H_{ext}$ for $\theta_H = 89°$ (red circles). Black circles are the extracted Gilbert damping, which is independent of $H_{ext}$. The black dotted line shows the average of this extracted damping; (d) and (e) depict similar plots of $1/\tau$ and damping constants for $\theta_H = 76°$. Error bars in (b) through (e) come from the uncertainty in the mathematical fitting.

The inverse relaxation time ($1/\tau$) should also have a minimum value near $H_{k,\text{eff}}$ for $\theta_H = 89°$ if the damping was purely from Gilbert damping [as shown by the solid lines in Figs. 4(b) and 4(d)]; however, the measured data do not follow this trend. Adding the inhomogeneous term [dotted lines in Figs. 4(b) and 4(d)] more accurately describes the field dependence of the measured $1/\tau$ [open symbols in Figs. 4(b) and 4(d)]. It should be noted that the dip of the predicted $1/\tau$ occurs when the frequency derivative term in Eq. (5) approaches zero; however, this is not captured by the measurement due to the finite interval over which we vary $H_{\text{ext}}$. Figures 4(c) and 4(e) depict the field-dependent effective damping ($\alpha_{\text{eff}}$) and the Gilbert damping ($\alpha$) as the intrinsic material's property obtained from fitting the measured $1/\tau$.

With the knowledge that the value of $\alpha$ extracted with this method is the intrinsic material property, we repeat this data reduction technique for the annealed W/CoFeB/MgO samples discussed in Fig. 1. The symbols in Fig. 5 represent the resonance frequency and damping constants (both effective damping and Gilbert damping) for all samples measured at $\theta_H \approx 90°$. The fittings for the resonance frequency [red lines, from Eq. (1)] are also shown to demonstrate the good agreement between our TR-MOKE measurement and theoretical prediction. The uncertainties of $f$, $\tau$, and $H_{k,\text{eff}}$ are calculated from the least-squares fitting uncertainty and the uncertainty of measuring $H_{\text{ext}}$ with the Hall sensor.

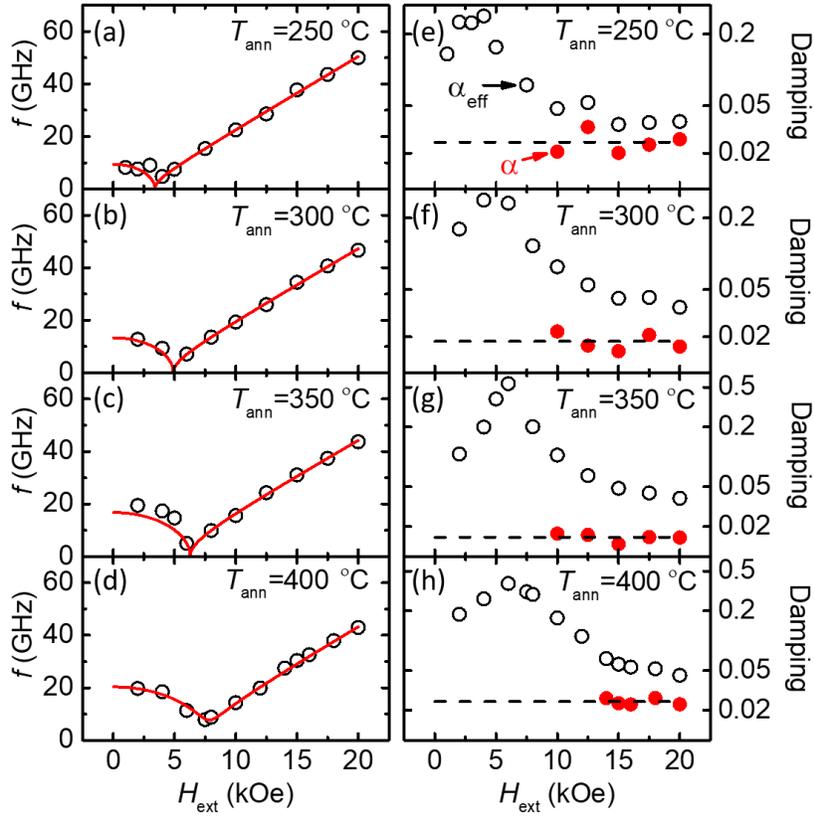

**Figure 5**. Results for $f$ (a-d) and $\alpha_{\text{eff}}$ (e-h, on a log scale) for individual samples. For comparison, the Gilbert damping constant $\alpha$ is also plotted by subtracting the inhomogeneous terms from $\alpha_{\text{eff}}$. The dashed line in (e-h) indicates the average $\alpha$. All samples are measured at $\theta_H = 90°$ except for the 400 °C sample ($\theta_H = 89°$).

The summary of the anisotropy and damping measured via TR-MOKE is shown in Fig. 6. Figure 6(a) plots $H_{k,\text{eff}}$ obtained from VSM (black open circles) and TR-MOKE (blue open squares), both of which exhibit a monotonic increasing trend as $T_{\text{ann}}$ becomes higher. Discrepancies in $H_{k,\text{eff}}$ from these two methods can be attributed to the difference in the size of the probing region, which is highly localized in TR-MOKE but sample-averaged in VSM. Since $H_{k,\text{eff}}$ determined from TR-MOKE is obtained from fitting the measured frequency for a localized region, we expect these values more consistently describe the magnetization precession than those obtained from VSM. The increase in $H_{k,\text{eff}}$ with $T_{\text{ann}}$ can be partially attributed to the crystallization of the CoFeB layer [32]. For temperatures higher than 350 °C, this increasing trend of $H_{k,eff}$ begins to lessen, presumably due to the diffusion of W atoms into the CoFeB layer, which is more pronounced at higher $T_{ann}$. The W diffusion process is also responsible for the decrease in $M_s$ of the CoFeB layer as $T_{ann}$ increases [Fig. 1(e)]. Subsequently, the decrease in $M_s$ leads to a further-reduced demagnetizing energy and thus a larger $H_{k,eff}$.

Similar observation of $M_s$ has been reported in literature for Ta/CoFeB/MgO PMA structures and attributed to the growth of a dead layer at the heavy metal/CoFeB interface [1]. Figure 6(b) summarizes $t_{dead}$ as a function of $T_{ann}$ with $t_{dead}$ increasing from 0.17 to 0.53 nm as $T_{ann}$ changes from 250 to 400 °C, as discussed in Section II.

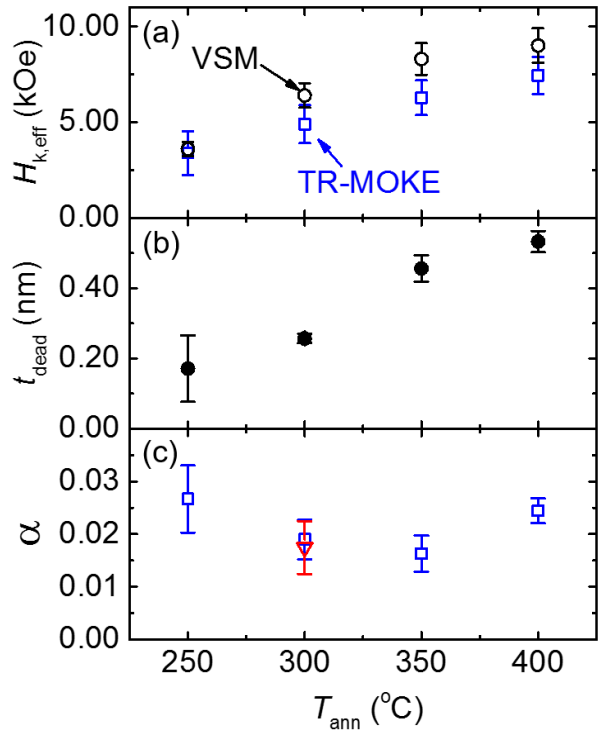

**Figure 6.** Summary of the magnetic properties of W-seeded CoFeB as a function of $T_{ann}$. (a) The dependence of $H_{k,eff}$ on $T_{ann}$ obtained from both the VSM (black open circles) and TR-MOKE fitting (blue open squares). (b) The dependence of dead-layer thickness on $T_{ann}$. (c) Damping constants as a function of $T_{ann}$. The minimum damping constant of $\alpha$ = 0.016 occurs at 350°C. The values for the all samples are obtained from measurements at $\theta_H$ = ~90°. For comparison, $\alpha$ of the reference Ta/CoFeB/MgO PMA sample annealed at 300 °C is also shown as a red triangle in (c).

Figure 6(c) depicts the dependence of $\alpha$ on $T_{\text{ann}}$, which first decreases with $T_{\text{ann}}$, reaches a minimum of 0.016 at 350 °C, and then increases as $T_{\text{ann}}$ rises to 400 °C. Similar trends have been observed for Ta/CoFeB/MgO previously (minimum $\alpha$ at $T_{\text{ann}} = 300$ °C) [32]. We speculate that this dependence of damping on $T_{\text{ann}}$ is due to two competing effects: (1) the increase in crystallization in the CoFeB layer with $T_{\text{ann}}$ which reduces the damping, and (2) the growth of a dead layer, which results from the diffusion of W and B atoms and is prominent at higher $T_{\text{ann}}$. At $T_{\text{ann}} = 400$ °C, the dead-layer formation leads to a larger damping presumably due to an increase in scattering sites (diffused atoms) that contribute to spin-flip events, as described by the Elliot-Yafet relaxation mechanisms [17]. The observation that our W-seeded samples still sustain excellent PMA properties at $T_{\text{ann}} = 400$ °C confirms their enhanced thermal stability, compared with Ta/CoFeB/MgO stacks which fail at $T_{\text{ann}} = 350$ °C or higher.

The damping constants are comparable for the W/CoFeB/MgO and Ta/CoFeB/MgO films annealed at 300 °C, both of which are higher than that of the W/CoFeB/MgO PMA with the optimal $T_{\text{ann}}$ of 350 °C. Nevertheless, our work focuses on the enhanced thermal stability of W-seeded CoFeB PMA films while still maintaining a relatively low damping constant. Such an advantage enables W-seeded CoFeB layers to be viable and promising alternatives to Ta/CoFeB/MgO, which is currently widely used in spintronic devices.

## V.   CONCLUSION

In summary, we deposit a series of W-seeded CoFeB PMA films with varying annealing temperatures up to 400 °C and conduct ultrafast all-optical TR-MOKE measurements to study their magnetization precession dynamics. The Gilbert damping, as an intrinsic material property, is proven to be independent of measurement conditions, such as the amplitudes and directions of the applied field. The damping constant varies with $T_{ann}$, first decreasing and then increasing, leading to a minimum of $\alpha = 0.016$ for the sample annealed at 350 °C. Due to the dead-layer growth, the damping constant slightly increases to $\alpha = 0.024$ at $T_{ann} = 400$ °C, comparable to the reference Ta/CoFeB/MgO PMA film annealed at 300°C, which demonstrates the improved enhanced thermal stability of W/CoFeB/MgO over the Ta/CoFeB/MgO structures. This strongly suggests the great potential of W/CoFeB/MgO PMA material systems for future spintronic device integration that requires materials to sustain a processing temperature as high as 400 °C.

**Acknowledgements**

This work is supported by C-SPIN (award #: 2013-MA-2381), one of six centers of STARnet, a Semiconductor Research Corporation program, sponsored by MARCO and DARPA. The authors would like to thank Prof. Paul Crowell and Dr. Changjiang Liu for valuable discussions.

**Supplemental Materials Available:** Complete description of the TR-MOKE measurement method, the angular dependent result summary, and the description for determining the interface anisotropy.

# References

[1] S. Ikeda, K. Miura, H. Yamamoto, K. Mizunuma, H. D. Gan, M. Endo, S. Kanai, J. Hayakawa, F. Matsukura, and H. Ohno, A perpendicular-anisotropy CoFeB-MgO magnetic tunnel junction, Nat. Mater. **9**, 721 (2010).

[2] H. Sato, E. C. I. Enobio, M. Yamanouchi, S. Ikeda, S. Fukami, S. Kanai, F. Matsukura, and H. Ohno, Properties of magnetic tunnel junctions with a MgO/CoFeB/Ta/CoFeB/MgO recording structure down to junction diameter of 11 nm, Appl. Phys. Lett. **105**, 062403 (2014).

[3] J. Z. Sun, S. L. Brown, W. Chen, E. A. Delenia, M. C. Gaidis, J. Harms, G. Hu, X. Jiang, R. Kilaru, W. Kula, G. Lauer, L. Q. Liu, S. Murthy, J. Nowak, E. J. O'Sullivan, S. S. P. Parkin, R. P. Robertazzi, P. M. Rice, G. Sandhu, T. Topuria, and D. C. Worledge, Spin-torque switching efficiency in CoFeB-MgO based tunnel junctions, Phys. Rev. B **88**, 104426 (2013).

[4] J. Zhu, J. A. Katine, G. E. Rowlands, Y.-J. Chen, Z. Duan, J. G. Alzate, P. Upadhyaya, J. Langer, P. K. Amiri, K. L. Wang, and I. N. Krivorotov, Voltage-Induced Ferromagnetic Resonance in Magnetic Tunnel Junctions, Phys. Rev. Lett. **108**, 197203 (2012).

[5] W. G. Wang, M. G. Li, S. Hageman, and C. L. Chien, Electric-field-assisted switching in magnetic tunnel junctions, Nat. Mater. **11**, 64 (2012).

[6] S. Kanai, M. Yamanouchi, S. Ikeda, Y. Nakatani, F. Matsukura, and H. Ohno, Electric field-induced magnetization reversal in a perpendicular-anisotropy CoFeB-MgO magnetic tunnel junction, Appl. Phys. Lett. **101**, 122403 (2012).

[7] G. Q. Yu, P. Upadhyaya, Y. B. Fan, J. G. Alzate, W. J. Jiang, K. L. Wong, S. Takei, S. A. Bender, L. T. Chang, Y. Jiang, M. R. Lang, J. S. Tang, Y. Wang, Y. Tserkovnyak, P. K. Amiri, and K. L. Wang, Switching of perpendicular magnetization by spin-orbit torques in the absence of external magnetic fields, Nat. Nanotechnol. **9**, 548 (2014).

[8] D. Bhowmik, L. You, and S. Salahuddin, Spin Hall effect clocking of nanomagnetic logic without a magnetic field, Nat. Nanotechnol. **9**, 59 (2014).

[9] L. Q. Liu, C. F. Pai, Y. Li, H. W. Tseng, D. C. Ralph, and R. A. Buhrman, Spin-Torque Switching with the Giant Spin Hall Effect of Tantalum, Science **336**, 555 (2012).

[10] W. J. Jiang, P. Upadhyaya, W. Zhang, G. Q. Yu, M. B. Jungfleisch, F. Y. Fradin, J. E. Pearson, Y. Tserkovnyak, K. L. Wang, O. Heinonen, S. G. E. te Velthuis, and A. Hoffmann, Blowing magnetic skyrmion bubbles, Science **349**, 283 (2015).

[11] G. Yu, P. Upadhyaya, Q. Shao, H. Wu, G. Yin, X. Li, C. He, W. Jiang, X. Han, P. K. Amiri, and K. L. Wang, Room-Temperature Skyrmion Shift Device for Memory Application, Nano Letters **17**, 261 (2017).

[12] Z. T. Diao, Z. J. Li, S. Y. Wang, Y. F. Ding, A. Panchula, E. Chen, L. C. Wang, and Y. M. Huai, Spin-transfer torque switching in magnetic tunnel junctions and spin-transfer torque random access memory, J. Phys.: Condens. Matter **19**, 165209 (2007).

[13] S. Manipatruni, D. E. Nikonov, and I. A. Young, Energy-delay performance of giant spin Hall effect switching for dense magnetic memory, Appl. Phys. Express **7**, 103001 (2014).

[14] A. J. Annunziata, P. L. Trouilloud, S. Bandiera, S. L. Brown, E. Gapihan, E. J. O'Sullivan, and D. C. Worledge, Materials investigation for thermally-assisted magnetic random access memory robust against 400 °C temperatures, J. Appl. Phys. **117**, 17B739 (2015).

[15] N. Miyakawa, D. C. Worledge, and K. Kita, Impact of Ta Diffusion on the Perpendicular Magnetic Anisotropy of Ta/CoFeB/MgO, IEEE Magn. Lett. **4**, 1000104 (2013).

[16] S. Yang, J. Lee, G. An, J. Kim, W. Chung, and J. Hong, Thermally stable perpendicular magnetic anisotropy features of Ta/TaO$_x$/Ta/CoFeB/MgO/W stacks via TaO$_x$ underlayer insertion, J. Appl. Phys. **116**, 113902 (2014).
[17] R. J. Elliott, Theory of the Effect of Spin-Orbit Coupling on Magnetic Resonance in Some Semiconductors, Phys. Rev. **96**, 266 (1954).
[18] H. Suhl, Theory of the magnetic damping constant, IEEE Trans. Magn. **34**, 1834 (1998).
[19] S. Ikeda, J. Hayakawa, Y. Ashizawa, Y. M. Lee, K. Miura, H. Hasegawa, M. Tsunoda, F. Matsukura, and H. Ohno, Tunnel magnetoresistance of 604% at 300K by suppression of Ta diffusion in CoFeB/MgO/CoFeB pseudo-spin-valves annealed at high temperature, Appl. Phys. Lett. **93**, 082508 (2008).
[20] T. Liu, Y. Zhang, J. W. Cai, and H. Y. Pan, Thermally robust Mo/CoFeB/MgO trilayers with strong perpendicular magnetic anisotropy, Sci. Rep. **4**, 5895 (2014).
[21] H. Almasi, D. R. Hickey, T. Newhouse-Illige, M. Xu, M. R. Rosales, S. Nahar, J. T. Held, K. A. Mkhoyan, and W. G. Wang, Enhanced tunneling magnetoresistance and perpendicular magnetic anisotropy in Mo/CoFeB/MgO magnetic tunnel junctions, Appl. Phys. Lett. **106**, 182406 (2015).
[22] S. Cho, S.-h. C. Baek, K.-D. Lee, Y. Jo, and B.-G. Park, Large spin Hall magnetoresistance and its correlation to the spin-orbit torque in W/CoFeB/MgO structures, Sci. Rep. **5**, 14668 (2015).
[23] C. F. Pai, L. Q. Liu, Y. Li, H. W. Tseng, D. C. Ralph, and R. A. Buhrman, Spin transfer torque devices utilizing the giant spin Hall effect of tungsten, Appl. Phys. Lett. **101**, 122404 (2012).
[24] C. He, A. Navabi, Q. Shao, G. Yu, D. Wu, W. Zhu, C. Zheng, X. Li, Q. L. He, S. A. Razavi, K. L. Wong, Z. Zhang, P. K. Amiri, and K. L. Wang, Spin-torque ferromagnetic resonance measurements utilizing spin Hall magnetoresistance in W/Co$_{40}$Fe$_{40}$B$_{20}$/MgO structures, Appl. Phys. Lett. **109**, 202404 (2016).
[25] M. van Kampen, C. Jozsa, J. T. Kohlhepp, P. LeClair, L. Lagae, W. J. M. de Jonge, and B. Koopmans, All-Optical Probe of Coherent Spin Waves, Phys. Rev. Lett. **88**, 227201 (2002).
[26] S. Mizukami, D. Watanabe, T. Kubota, X. Zhang, H. Naganuma, M. Oogane, Y. Ando, and T. Miyazaki, Laser-Induced Fast Magnetization Precession and Gilbert Damping for CoCrPt Alloy Thin Films with Perpendicular Magnetic Anisotropy, Appl. Phys. Express **3**, 123001 (2010).
[27] S. Mizukami, F. Wu, A. Sakuma, J. Walowski, D. Watanabe, T. Kubota, X. Zhang, H. Naganuma, M. Oogane, Y. Ando, and T. Miyazaki, Long-Lived Ultrafast Spin Precession in Manganese Alloys Films with a Large Perpendicular Magnetic Anisotropy, Phys. Rev. Lett. **106**, 117201 (2011).
[28] S. Iihama, S. Mizukami, N. Inami, T. Hiratsuka, G. Kim, H. Naganuma, M. Oogane, T. Miyazaki, and A. Yasuo, Observation of Precessional Magnetization Dynamics in L1$_0$-FePt Thin Films with Different L1$_0$ Order Parameter Values, Jpn. J. Appl. Phys. **52**, 073002 (2013).
[29] J. Zhu, X. Wu, D. M. Lattery, W. Zheng, and X. Wang, The Ultrafast Laser Pump-probe Technique for Thermal Characterization of Materials With Micro/nanostructures, Nanoscale Microscale Thermophys. Eng., (2017). [published online]
[30] C. Y. You and S. C. Shin, Generalized analytic formulae for magneto-optical Kerr effects, J. Appl. Phys. **84**, 541 (1998).
[31] P. Krivosik, N. Mo, S. Kalarickal, and C. E. Patton, Hamiltonian formalism for two magnon scattering microwave relaxation: Theory and applications, J. Appl. Phys. **101**, 083901 (2007).

[32] S. Iihama, S. Mizukami, H. Naganuma, M. Oogane, Y. Ando, and T. Miyazaki, Gilbert damping constants of Ta/CoFeB/MgO(Ta) thin films measured by optical detection of precessional magnetization dynamics, Phys. Rev. B **89**, 174416 (2014).
[33] C. Kittel, On the Theory of Ferromagnetic Resonance Absorption, Phys. Rev. **73**, 155 (1948).
[34] L. Ma, S. F. Li, P. He, W. J. Fan, X. G. Xu, Y. Jiang, T. S. Lai, F. L. Chen, and S. M. Zhou, Tunable magnetization dynamics in disordered FePdPt ternary alloys: Effects of spin orbit coupling, J. Appl. Phys. **116**, 113908 (2014).
[35] S. Mizukami, Fast Magnetization Precession and Damping for Magnetic Films with High Perpendicular Magnetic Anisotropy, J. Magn. Soc. Jpn. **39**, 1 (2015).
[36] H. Suhl, Ferromagnetic Resonance in Nickel Ferrite Between One and Two Kilomegacycles, Phys. Rev. **97**, 555 (1955).
[37] D. S. Wang, S. Y. Lai, T. Y. Lin, C. W. Chien, D. Ellsworth, L. W. Wang, J. W. Liao, L. Lu, Y. H. Wang, M. Z. Wu, and C. H. Lai, High thermal stability and low Gilbert damping constant of CoFeB/MgO bilayer with perpendicular magnetic anisotropy by Al capping and rapid thermal annealing, Appl. Phys. Lett. **104**, 142402 (2014).
[38] A. Capua, S. H. Yang, T. Phung, and S. S. P. Parkin, Determination of intrinsic damping of perpendicularly magnetized ultrathin films from time-resolved precessional magnetization measurements, Phys. Rev. B **92**, 224402, 224402 (2015).